# Activation of Phosphorene-like Two-dimensional GeSe for Efficient Electrocatalytic Nitrogen Reduction via States Filtering of Ru

Zheng Shu, [a] Yongqing Cai[b*]

[a]Center for Data Science, Institute of Collaborative Innovation, University of Macau, Taipa, Macau, China

[b]Joint Key Laboratory of the Ministry of Education, Institute of Applied Physics and Materials Engineering, University of Macau, Taipa, Macau, China

Email: yongqingcai@um.edu.mo

## Abstract

Nitrogen reduction reaction (NRR) which converts nitrogen ($N_2$) to ammonia ($NH_3$) normally requires harsh conditions to break the bound nitrogen bond. Herein, via first-principles calculation we reveal that a superior NRR catalytic activity could be obtained through anchoring atomic catalyst above a phosphorene-like puckering surface of germanium selenide (GeSe). Through examining the single- and double- atoms (B, Fe, W, Mo and Ru) decorated on GeSe, we find that its rippled structure allows an intimate contact between the deposited species and the GeSe which significantly promotes the states hybridization. Amongst the various atomic catalyst, we predict that the Ru dimer decorated GeSe monolayer ($Ru_2$@GeSe) has superior catalytic activity for the $N_2$ fixation and reduction. Through examining the three NRR pathways (distal, alternating and enzymatic), the distal and enzymatic pathway is both the thermodynamically favorable with the maximum Gibbs free energy change ($\Delta G_{MAX}$) of 0.25 and 0.26 eV, respectively. Such a superior activity could be attributed to the filtered states of GeSe by Ru dimer which leads to the effective activation of the adsorbed $N_2$ bond. As an efficient near-infrared absorber of GeSe, the Ru mediated hybridization of GeSe-Ru-$N_2$ complex enables an in-gap state which further broadens the absorbing window, rendering for a broadband solar absorption and possible photocatalysis.

## 1. Introduction

Ammonia ($NH_3$) is an extreme important chemical which is not only a carbon-free energy carrier but also plays a wide usage in agricultural and industrial applications[1-3]. In general, the $N_2$ reduction reaction (NRR) into ammonia ($NH_3$) is synthesized through traditional Haber-Bosch process that combines nitrogen ($N_2$) and hydrogen ($H_2$) molecules into ammonia ($NH_3$) under high pressure and temperature[4]. Due to the large nitrogen binding energy of 962 kJ·mol$^{-1}$, the traditional Haber-Bosch approach requires harsh thermodynamic conditions with significant amounts of energy uptake each year (up to 1% share of total world`s energy supply[1]). Identifying alternative approaches with cost-effectiveness, sustainable and environmental friendly nature for NRR, in particularly via renewable energy, is intensively explored. However, finding efficient approaches to fix and activate inert triple bond of $N_2$ still remains a challenge.

Fixation and activation of inert triple bond of $N_2$ are intrinsically difficult owing to the inert N-N triple bond[5-12]. To this end, the natural process of $N_2$ fixation via nitrogenase enzyme at ambient conditions provides hints for the nitrogen bonding dissociation which triggers great interests for efficient kinetically nitrogen reduction[6]. Electrocatalytic NRR is proved to be an ideal and promising approach with utilization of the hydrogen renewably obtained from water[5]. As inspired by the FeMo functional center in nitrogenase enzyme[8], various non-metal catalyst i.e boron single atom catalyst[9-12], and transition metal (TM)-based catalysts[13-16] have been examined for the conversion of dinitrogen into ammonia. Liu *et al.*[11] showed B single atom anchored on graphene exhibits relatively low $\Delta G_{MAX}$ with 0.31 eV. One W-BP (W atoms substitute the surface P atoms of BP) was presented[13] with onset potentials of 0.42 eV. Wu *et al.*[14] discovered Mo-doped $Fe_2P$ can promote the NRR with $\Delta G_{MAX}$ of 0.30 eV. And Yang *et al.*[53] proved Mo@$MoS_2$ has overpotential of 0.28 V by distal pathway. However, for a thermodynamical activation of nitrogen bond at low temperature and moderate pressure conditions, a proper match of the evanescent states of nano TM species and the extending states of the host for efficient injection of carriers to the frontier orbitals of $N_2$ is critical.

In this work, we demonstrate a new-type of NRR catalyst based on monochalcogenides supported adatoms. Group-IV chalcogenides MX (M = Ge, Sn; X = S, Se), which are isostructural analogue of phosphorene, have attracted much attention due to their potential application in nano-electronic device, resistive switching devices, electrochemical memory cells and advanced catalysts[17-24]. Among these MX 2D monolayers, GeSe is the only compound which has direct band structure[25,26], even though there is a very small energy difference between direct and indirect band gap[25]. Unlike phosphorene, GeSe is much more stable[26-28]. There are many relevant researches that prove potential catalytic activity of GeSe monolayer[29-32]. However, the NRR activity of GeSe monolayer still remains unknown. Herein, we systematically explored the performance of single and double atomic (B, Fe, W, Mo and Ru) decorated GeSe monolayer ($D_x$@GeSe) for NRR. We demonstrate that Ru dimer on GeSe monolayer could possess promising superior catalytic performance with lowest $\Delta G_{MAX\text{-}Enzymatic}$ = 0.26 eV, $\Delta G_{MAX\text{-}Distal}$ = 0.25 eV and a rather low overpotential ($\eta_{Enzymatic}$ = 0.10 eV and $\eta_{Distal}$ = 0.09 eV). The high catalytic NRR activity arises from synergistic effect of Ru and GeSe which forms hybridized state that can effectively activate the supported $N_2$ bond, allowing associative splitting via sequential hydrogenation.

## 2. Computational methodology

All structural optimizations and energy calculations based on spin-polarized density functional theory (DFT) were implemented with the Vienna ab initio simulation package (VASP)[33]. Generalized gradient approximation (GGA) in the Perdew--Burke--Ernzerhof (PBE) form was adopted to describe electron exchange and correlation effects. We used a supercell (4 × 4 × 1) of GeSe monolayer structure with 64 atoms referred to our previous work[20] as the substrate for catalysts and the cutoff energy of 450 eV. We adopted Monkhorst-Pack k point mesh of the 3 × 3 × 1 grid for sampling points in the first Brillouin zone of the supercell. A vacuum layer with thickness of 20 Å was chosen to avoid interlayer interaction of image layers under the periodic boundary condition. All crystal geometric configurations reached required accuracy with the system energies and residual forces on each atom less than $1 \times 10^{-5}$ eV and 0.01 eV $Å^{-1}$, respectively. Grimme's semiempirical DFT-D3 method[34] was utilized to address the long-range van der Waals (vdW) interaction. The charge transfer was calculated via Bader charge analysis method. Furthermore, we used ab initio molecular dynamics (AIMD) simulations to valid the thermodynamic stability of $Ru_2$@GeSe.

In NRR catalytic reaction, the change of Gibbs free energy ($\Delta G$) of each elemental step is the figure of merit for catalytic activity. Our calculations of $\Delta G$ are followed by computational hydrogen electrode model presented by Nørskov[35] *et al.*, which is formulated as:

$$\Delta G = \Delta E + \Delta E_{ZPE} - T\Delta S + eU + \Delta GpH \qquad (1)$$

where $\Delta E$ represents energy difference between products and reactants of each elementary step in the NRR; $\Delta E_{ZPE}$ and $\Delta S$ are the changes of zero-point energy (ZPE) and entropy, respectively, which can be calculated from vibrational frequencies, and $T$ is the temperature which is set to 298.15 K in our work. The calculations of $\Delta E_{ZPE}$ and $T\Delta S$ are based on the following equations:

$$E_{ZPE} = \frac{1}{2}\sum_i h\,\nu_i \qquad (2)$$

$$-TS = k_B T \sum_i \ln\left(1 - e^{-\frac{h\nu_i}{k_B T}}\right) - \sum_i h\nu_i \left(\frac{1}{e^{\frac{h\nu_i}{k_B T}} - 1}\right) \qquad (3)$$

where $k_B$, $h$ and $\nu_i$ denote Boltzmann constant, Planck constant and vibrational frequencies of mode $i$, respectively. For molecular $H_2$, $NH_3$ and $N_2$, standard values of $E_{ZPE}$ and $T\Delta S$ were used in our work, which are listed in Table S1. $\Delta GpH = k_B T \times$ pH × In10 is the effect of pH on $\Delta G$, whereas pH is set to 0 under the standard condition. The free energy correction of $eU$ is the item which is related to applied electrode potential ($U$), whereas e is the number of transferred electrons. The potential-determining step for the whole NRR process is the elementary step which has the most positive free energy change ($\Delta G_{MAX}$). The onset potential $U_{onset}$ = -$\Delta G_{MAX}$/e. The overpotential η is determined by η = $U_{equillibrium}$ - $U_{onset}$, where $U_{equillibrium}$ of NRR is -0.16 V. The post-process of optical properties of GeSe calculated by VASP is implemented by VASPKIT[36] script. AIMD[35] were carried out to evaluate the thermodynamic stability in NRR condition with the presence of water up to 10 ps. Furthermore, we used nudges elastic band (NEB)[37] method to calculate the kinetic barriers for the limiting potential steps of each pathway. Although the solvent effect is proved to be negligible for NRR by many previous works[38-42], we still discuss it for thorough exploration by VASPsol[43,44] tool.

# 3. Results and Discussion

## 3.1 Preliminary screening of $D_x$@GeSe substrates for promising NRR

The $N_2$ free molecule contains lone-pair electrons and very bound covalent N-N bond. Potential NRR catalysts would have localized empty levels to host those transferred electrons from bonding states (σ bond) of $N_2$, while simultaneously donate electrons to the empty anti-bonding states (π*) of $N_2$. Such a back-donation mechanism is a necessity for an efficient dissociative adsorption of $N_2$. Herein we screened several structures with anchoring single and double dopants above GeSe ($D_x$@GeSe, $x$=1 and 2 representing single and double dopants, respectively) for potential NRR (D = Ru, Fe, W, Mo and B). Those dopants were selected due to the presence of empty d orbitals, and half-filled $sp^3$ hybridized states in boron complex[10]. In addition, the rich empty d-states in Ru and Fe, and $sp^3$ hybridization of B would act as an exchange center allowing the tunneling of the excited carriers of GeSe (intrinsically p-type) into the bonding/antibonding states of $N_2$. Moreover, Mo and W atom decorated on 2D materials also present superior catalytic activity[45-47]. In the NRR, the initial $N_2$ molecule can undergo two adsorption possibilities—the so-called end-on adsorption where only one end of $N_2$ molecule is bonded and side-on adsorption where two ends of $N_2$ are bonded. It should be noted that either approach has a different hydrogenation process. For the end-on configuration there are two reaction pathways for hydrogenation: distal and alternating, while for the side-on configuration one reaction pathway, enzymatic route, is possible.

.

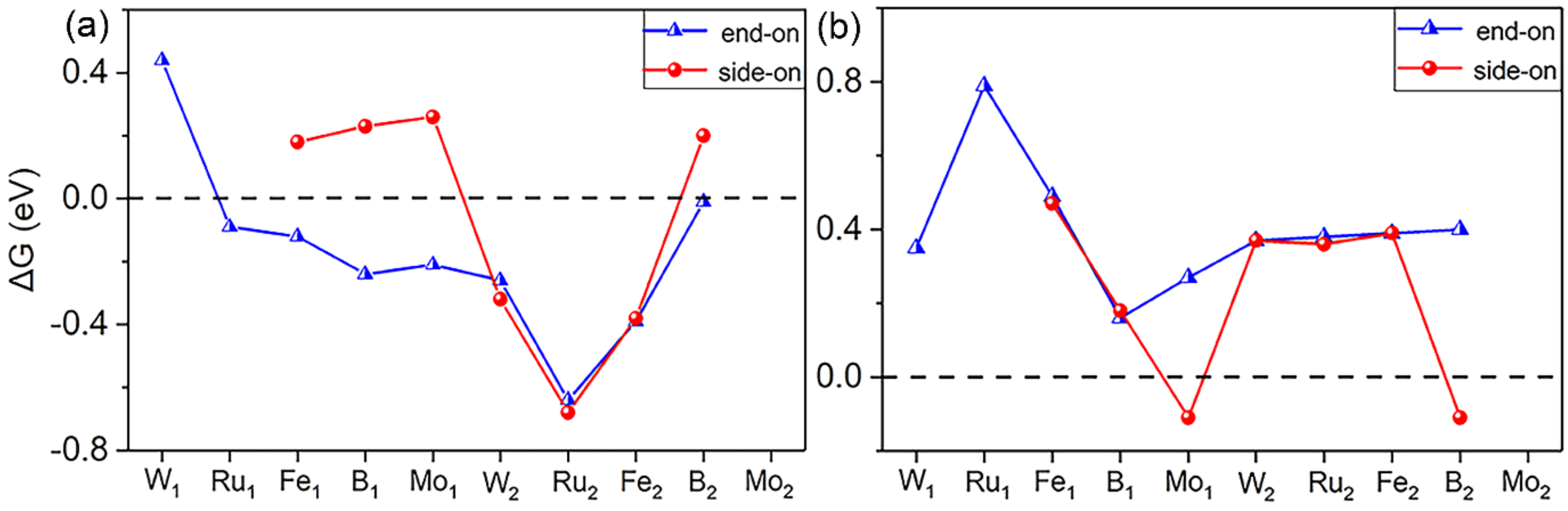

**Fig. 1** The variation of Gibbs free energy ($\Delta G$) by DFT calculation for (a) $N_2$ adsorption and (b) first-hydrogenation process on single and double atoms decorated (B, Fe, W, Mo and Ru) GeSe monolayer substrate.

To be as a potential NRR catalyst, it is a prerequisite of a chemical adsorption of the free $N_2$ gas molecule. The second factor governs the NRR would be an efficient capturing of the hydrogen above the adsorbed $N_2$ (denoted as $*N_2$). In particular, the first hydrogenation process, with forming $*N_2H$, is usually regarded as the potential-determining step (PDS)[48,49] of whole catalytic reaction. Based on these considerations, we evaluate the changes of Gibbs free energy (ΔG) for $N_2$ absorption and *$N_2H$ formation process on $D_x$@GeSe systems to identify which $D_x$@GeSe substrate may be a promising catalyst for the NRR. The calculated results of ΔG for the first two steps of NRR are shown in Fig. 1 (a) and (b), respectively.

We find that for single atom catalysts, $Ru_1$@GeSe and $W_1$@GeSe can't absorb $N_2$ via the side-on pathway. In contrast, other single atom catalysts can fix $N_2$ by both end-on and side-on pathways, but $B_1$@GeSe, $Fe_1$@GeSe and $Mo_1$@GeSe with side-on adsorption are endothermic with ΔG of 0.23 eV, 0.18 eV and 0.26 eV respectively, which means their $N_2$ adsorption process is not energetic favorable. Therefore $B_1$@GeSe, $Fe_1$@GeSe and $Mo_1$@GeSe are also not favorable for NRR due to the thermodynamically uphill step of the chemical adsorption of $N_2$. In this consideration, $D_1$@GeSe is not favorable for NRR. Next, we examine the $D_2$@GeSe systems ($B_2$@GeSe, $Fe_2$@GeSe, $W_2$@GeSe, $Mo_2$@GeSe and $Ru_2$@GeSe) in the following discussion. As we can see in Fig. 1 (a), $N_2$ adsorbed on $B_2$@GeSe is an endothermic process for side-on (ΔG = 0.20 eV) pathways. Thus, we also safely excluded the B dimer case. For $Fe_2$@GeSe, although $N_2$ adsorption process is a spontaneous reaction, the first hydrogenation process shown in Fig. 1 (b) requires a relative high energy input of 0.38 eV, thus also being ruled out. Like $Fe_2$@GeSe, the first hydrogenation process of $W_2$@GeSe requires 0.36 eV energy input by end-on pathway and 0.38 eV by side-on pathway, so we rule out this condition. Moreover, Mo dimer can't absorb $N_2$ although we try many different initial configurations for optimization. Interestingly, for Ru dimer case ($Ru_2$@GeSe), the $N_2$ adsorption is significantly exothermic and the Fig. 1 (a) shows the free energy change of side-on configuration ($\Delta G$ = -0.68 eV) is slightly lower than end-on configuration ($\Delta G$ = -0.64 eV). This implies that the side-on pathway for $N_2$ splitting would be more probable than the end-on pathway. It demonstrates that $Ru_2$@GeSe favors the side-on pathway over end-on pathway. Therefore, Ru dimer may be a promising atomic catalyst for NRR. The whole catalytic process on $Ru_2$@GeSe was further detailed examined in the subsequent sections.

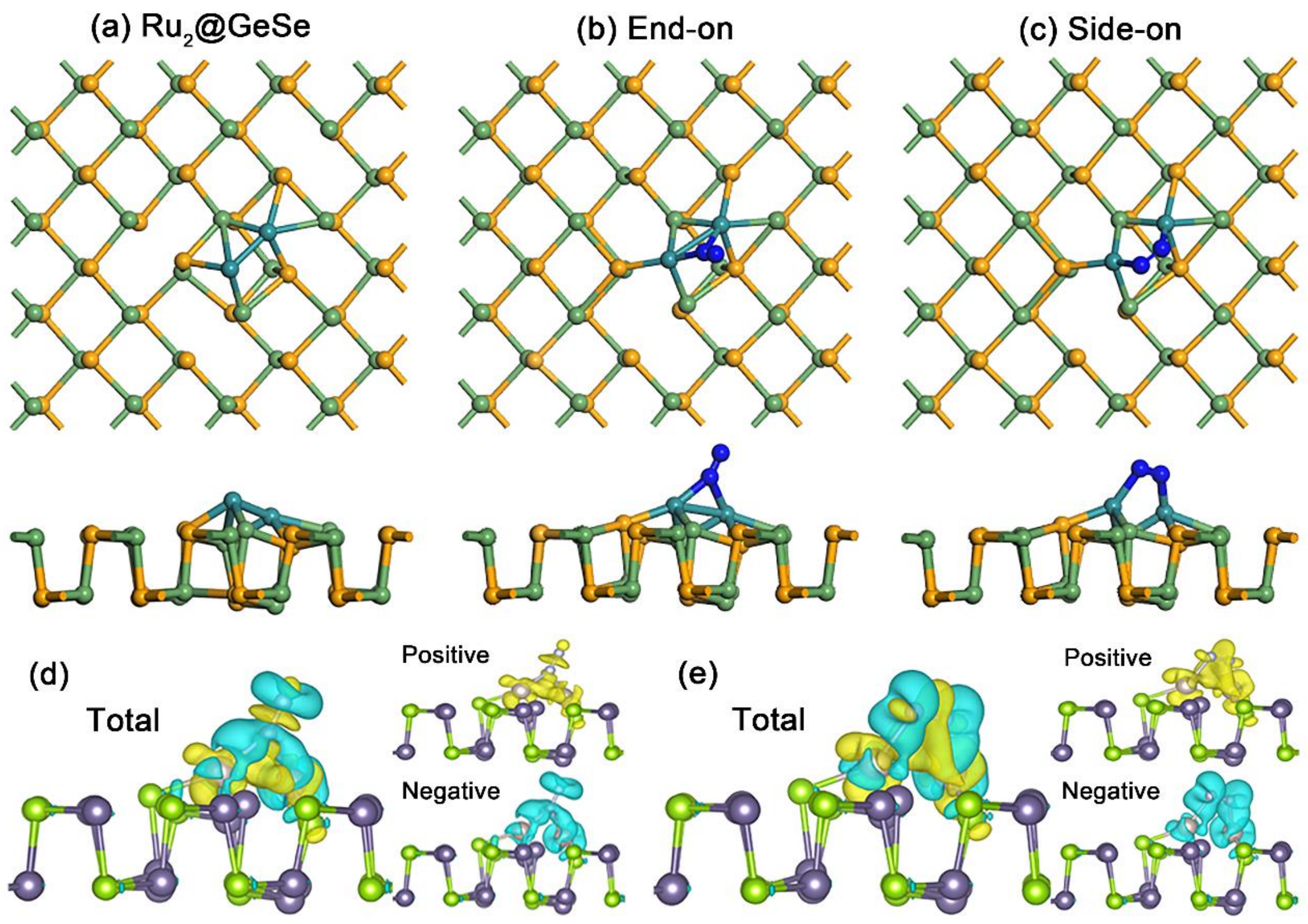

**Fig. 2** The top and side views of the optimized configurations of (a) $Ru_2$@GeSe substrate and the adsorption of $N_2$ via (b) end-on and (c) side-on patterns. The differential charge density of $Ru_2$@GeSe with the adsorption of $N_2$ through (d) end-on and (e) side-on patterns. The blue (yellow) color represents a loss (accumulation) of electrons. Green, yellow and blue atomic balls represent Ge, Se and N atoms, respectively.

### 3.2 Whole catalytic process of $N_2$ to $NH_3$

The GeSe surface adopts a puckered structure similar with phosphorene. The most energetic favorable binding geometry for Ru dimer is shown in Fig. 2 (a). The Ru dimer shows a strong chemical affinity toward GeSe with forming a flattened pattern above the sheet: one Ru atom is coordinated with one Ge atom and two Se atoms, whereas the other Ru atom is coordinated with one Se atom and two Ge atoms. The bond length of the Ru-Ru bond for the adsorbed Ru dimer is 2.40 Å.

Upon the uptake of the $N_2$ molecule, the N atoms are bonding with Ru atoms across two ridges and forming on a Ge-N-N-Ge-Ge ring-like structure above the GeSe surface, as shown in Fig. 2 (b) and (c). The bond length of the triple bond of $N_2$ was increased to 1.16 Å for end-on adsorption and 1.19 Å for side-on case compared with 1.12 Å for $N_2$ gas molecule, as indicative of effective activation of the $N_2$. This is consistent with a slightly stronger adsorption for the side-on case ($\Delta G$ = -0.68 eV) than the end-on geometry ($\Delta G$ = -0.64 eV), implying a correlation between the $N_2$ bond length and the energy release. The differential charge densities for the end-on and side-on $N_2$ were calculated and shown in Fig. 2(d) and (e), respectively. It can be seen that significant redistribution of electrons in the $N_2$ molecule occurs. An enhanced population (blue color) was found to show p-orbital antibonding states of the adsorbed $N_2$ molecule with characteristic lobes separated with zero nodal plane. Moreover, a depleted occupation (yellow color) appears in bonding states with covalent nature of nitrogen diatomic bond. Both enhanced antibonding bonding and weakened bonding states account for the elongation of the bond length of $N_2$ above $Ru_2$ which facilitates the dissociation of the homonuclear bond.

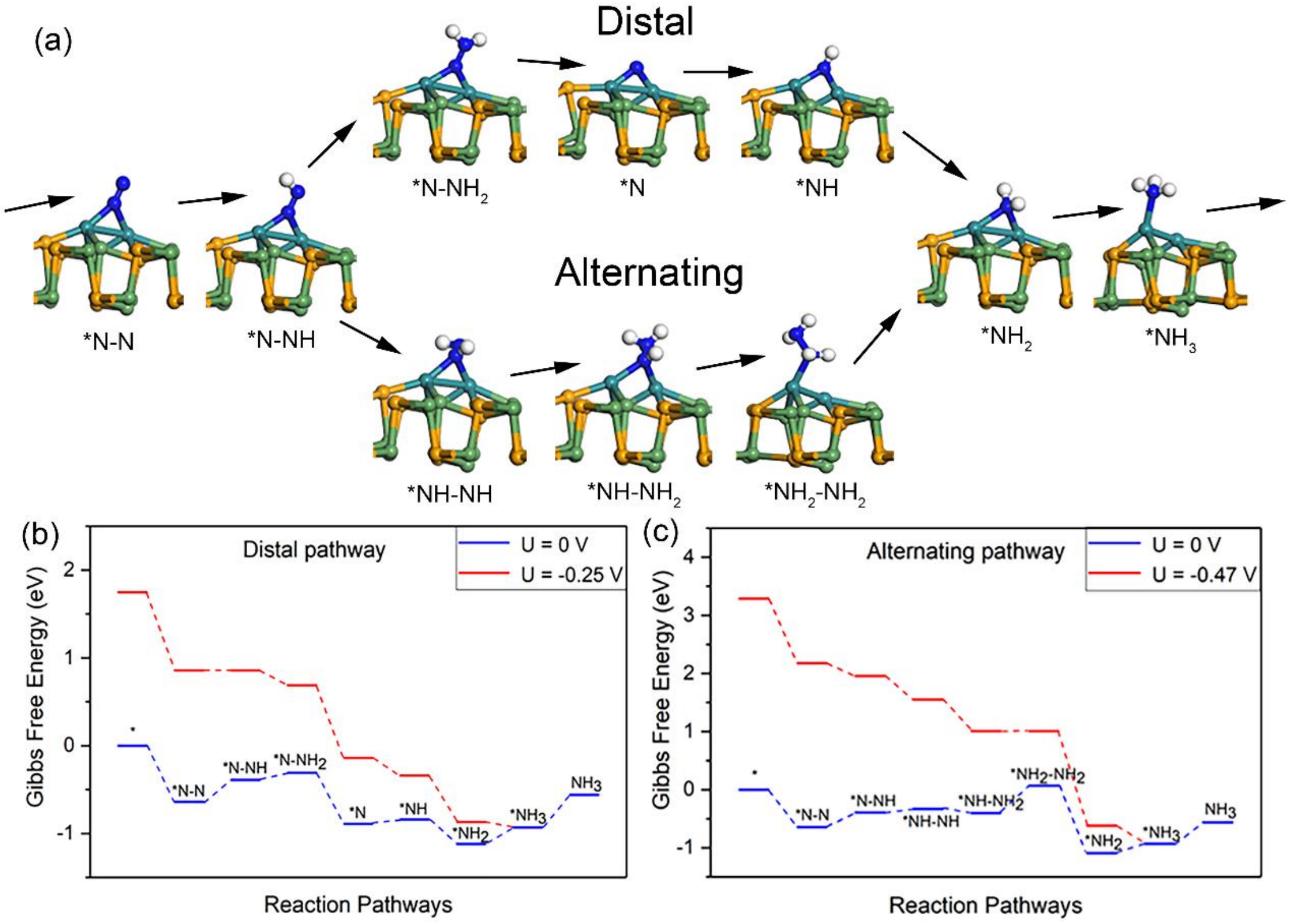

**Fig. 3** (a) Optimal structures of each reaction intermediate through distal and alternating pathways for $N_2$ reduction on $Ru_2$@GeSe; The Gibbs free energy of each elementary step for NRR through (b) distal and (c) alternating pathways at different potentials. The green, yellow, light blue, deep blue and white spheres represent germanium, selenium, ruthenium, nitrogen and hydrogen atoms, respectively.

We next examine energetics for the hydrogenation process with all possible reaction intermediates taken into account. For end-on adsorption, two routs of hydrogenation could be possible: the distal and alternating ways, which are shown in Fig. 3(a). Along the distal route, three hydrogens were sequentially bonded to the distal nitrogen atom followed by another three hydrogens adsorbing with the remaining nitrogen atom, while for the alternating pathway the hydrogen atoms were alternatingly bonded to the two nitrogen atoms. Fig. 3(b) and (c) display the diagrams for the change of Gibbs energy for distal and alternating pathways. For the distal way, our calculations demonstrate that the $N_2$ adsorption ($* + N_2 \rightarrow *N_2$), the release of the first $NH_3$ ($*N_2H_2 + H^+ + e^- \rightarrow *N + NH_3$) and ($*NH + H^+ + e^- \rightarrow *NH_2$) are exothermic with $\Delta G$ of -0.64, -0.58 and -0.28 eV, respectively. The potential-determining step is $*NH_2 \rightarrow *NH_3$ with the maximum change of Gibbs energy ($\Delta G_{MAX}$) is 0.25 eV, which suggests a high activity for NRR. For the alternating way, the three intermediate reactions, the $N_2$ adsorption ($*N_2 + H^+ + e^- \rightarrow *N_2H$), ($*NH\text{-}NH + H^+ + e^- \rightarrow *NH\text{-}NH_2$) and the release of the first $NH_3$ ($*NH_2\text{-}NH_2 + H^+ + e^- \rightarrow *NH_2 + NH_3$) are exothermic with $\Delta G$ of -0.64, -0.07, and -1.16 eV, respectively. Other processes are uphill reactions with the $\Delta G_{MAX}$ of 0.47 eV for *NH-

$NH_2 \rightarrow *NH_2\text{-}NH_2$. Although this value is slightly higher than that of distal way, it is still lower than bulk Ru catalyst[50,51]. For both approaches, the bond length of the nitrogen increases linearly with the gradual addition of hydrogen atoms (Fig. S1).

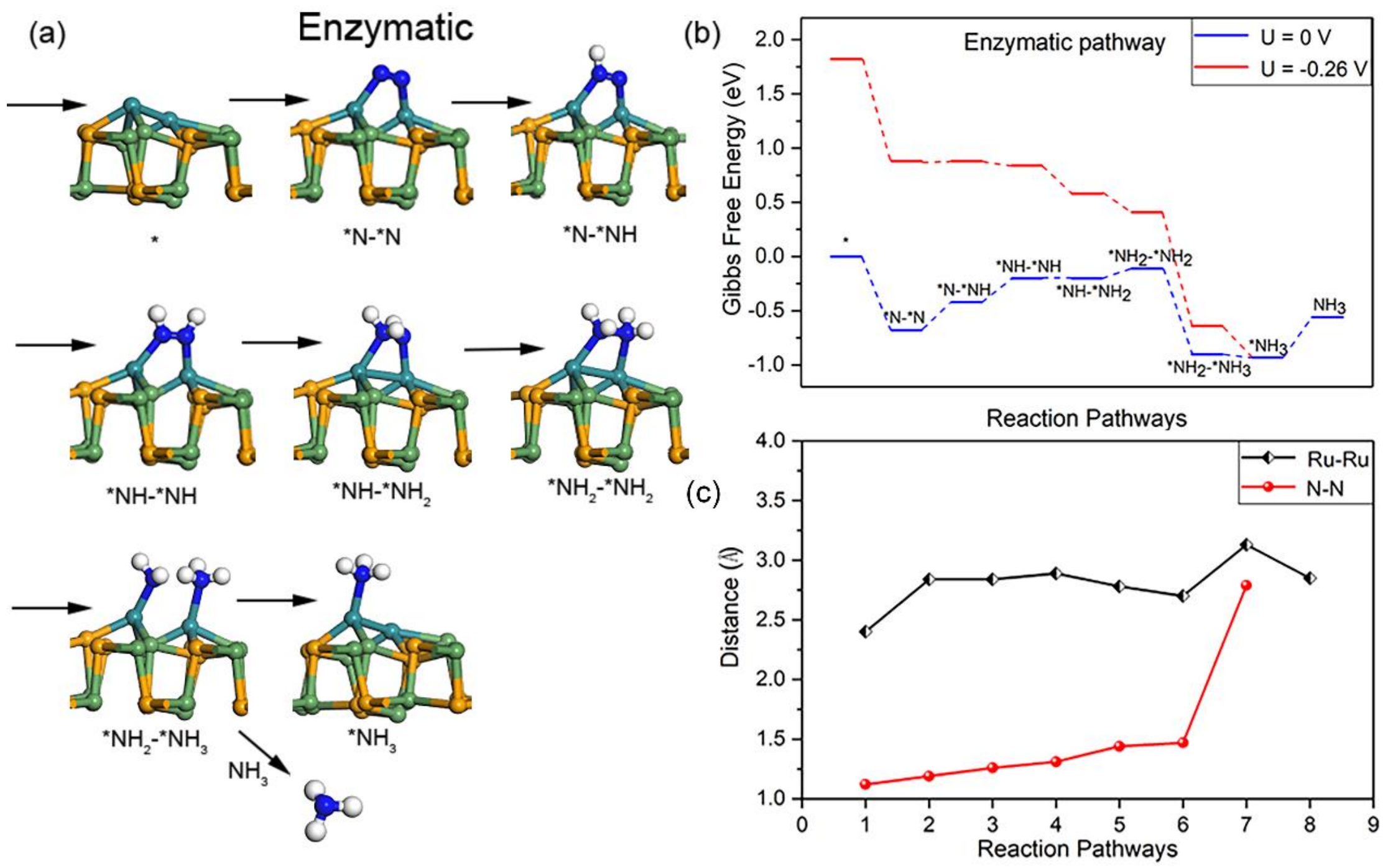

**Fig. 4** (a) Optimal structures of each reaction intermediate through enzymatic pathway for $N_2$ reduction on $Ru_2$@GeSe; (b) The Gibbs free energy of each elementary step for NRR through enzymatic pathways at different potentials. (c) The variations of bond length of Ru-Ru and N-N bonds via enzymatic pathway. The green, yellow, light blue, deep blue and white spheres represent germanium, selenium, ruthenium, nitrogen and hydrogen atoms, respectively.

For enzymatic pathway, as shown in Fig. 4 (a) and (b), there are three exothermic steps: $N_2$ adsorption ($*N_2 + H^+ + e^- \rightarrow *N_2H$)= -0.68 eV, ($*NH_2$-$*NH_2 + H^+ + e^- \rightarrow *NH_2$-$*NH_3$) = -0.79 eV and the first release of $NH_3$ for alternating way ($*NH_2$-$*NH_3 + H^+ + e^- \rightarrow *NH_3 + NH_3$) = -0.03 eV. The $G_{MAX}$ of enzymatic pathway is 0.26 eV which implies a low onset potential and a superior NRR activity. We used Bader charge analysis and discovered that 0.55 e charge was transferred to absorbed $N_2$ for the end-on configuration and 0.64 e charge for the side-on configuration. As shown in Fig. 4(c), with the continuous supply of hydrogen, the N-N bond is steadily elongated from 1.12 Å (adsorbed naked $N_2$) to 2.79 Å ($*NH_2$-$*NH_3$), indicating a dissociative splitting of the bound nitrogen bond. It should be noted that hydrogen evolution reaction (HER) is a competitive reaction compared with NRR, a high HER activity requires $\Delta G_{H*} \approx 0$ [29,52]. The $\Delta G(H*)$ for HER strongly deviates from this thermoneutral condition over 1 eV, implying a suppression of HER. Therefore, compared with the benchmarking Ru (0001) catalyst ($\Delta G_{MAX}$ = 1.08 eV)[50,51], we reveal that the $Ru_2$@GeSe catalyst has a much better performance for NRR with only $\Delta G_{MAX}$ = 0.26 eV through enzymatic process and 0.25 eV via the distal mechanism as shown above. Furthermore, the results also support that regarding the first hydrogenation process as the potential-determining step (PDS) is usually reliable.

**Table 1** The Calculated $\Delta G$ and $\Delta G_{sol}$ for each elementary step.

| Elementary Step | $\Delta G$/eV | $\Delta G_{sol}$/eV |
|---|---|---|
| $* + N_2 \rightarrow *N\text{-}*N$ | -0.68 | -0.69 |
| $*N\text{-}*N + H^+ + e^- \rightarrow *N\text{-}*NH$ | 0.26 | 0.22 |
| $*N\text{-}*NH + H^+ + e^- \rightarrow *NH\text{-}*NH$ | 0.22 | 0.15 |
| $*NH\text{-}*NH + H^+ + e^- \rightarrow *NH\text{-}*NH_2$ | 0 | -0.07 |
| $*NH\text{-}*NH_2 + H^+ + e^- \rightarrow *NH_2\text{-}*NH_2$ | 0.09 | -0.08 |
| $*NH_2\text{-}*NH_2 + H^+ + e^- \rightarrow *NH_2\text{-}*NH_3$ | -0.79 | -0.51 |
| $*NH_2\text{-}*NH_3 + H^+ + e^- \rightarrow *NH_3 + NH_3$ | -0.03 | -0.12 |
| $* + N_2 \rightarrow *N\text{-}N$ | -0.64 | -0.66 |
| $*N\text{-}N + H^+ + e^- \rightarrow *N\text{-}NH$ | 0.25 | 0.22 |
| $*N\text{-}NH + H^+ + e^- \rightarrow *N\text{-}NH_2$ | 0.08 | -0.01 |
| $*N\text{-}NH_2 + H^+ + e^- \rightarrow *N + NH_3$ | -0.58 | -0.46 |
| $*N + H^+ + e^- \rightarrow *NH$ | 0.05 | 0.04 |
| $*NH + H^+ + e^- \rightarrow *NH_2$ | -0.28 | -0.3 |
| $*NH_2 + H^+ + e^- \rightarrow *NH_3$ | 0.19 | 0.06 |
| $*N\text{-}NH + H^+ + e^- \rightarrow *NH\text{-}NH$ | 0.06 | 0.04 |
| $*NH\text{-}NH + H^+ + e^- \rightarrow *NH\text{-}NH_2$ | -0.07 | -0.22 |
| $*NH\text{-}NH_2 + H^+ + e^- \rightarrow *NH_2\text{-}NH_2$ | 0.47 | 0.5 |
| $*NH_2\text{-}NH_2 + H^+ + e^- \rightarrow *NH_2 + NH_3$ | -1.16 | -1.02 |

To explore NRR performance more thoroughly, we considered the solvent effect for the change of Gibbs free energy. We used VASPsol code which is an implicit solvation model for further evaluation of NRR on $Ru_2$@GeSe. The results showed that this correction is less than 0.2 eV. For distal and enzymatic pathway, the $\Delta G_{MAX}$ is reduced to 0.22 eV, thus the solvent here playing a minor role. The detailed free energy change is listed in Table 1.

Finally, nudges elastic band (NEB) method was used to evaluate kinetic barriers for $N_2$ absorption, the limiting potential steps of three pathways and $N_2$ desorption process. The results are plotted in Fig. S4. As we can see, the $N_2$ absorption process is always an exothermic process. The activation energies showed barriers is 0.14 eV, 0.14 eV and 0.05 eV for the limiting potential steps of enzymatic, distal and alternating pathways respectively, which are very low barriers at room temperature.

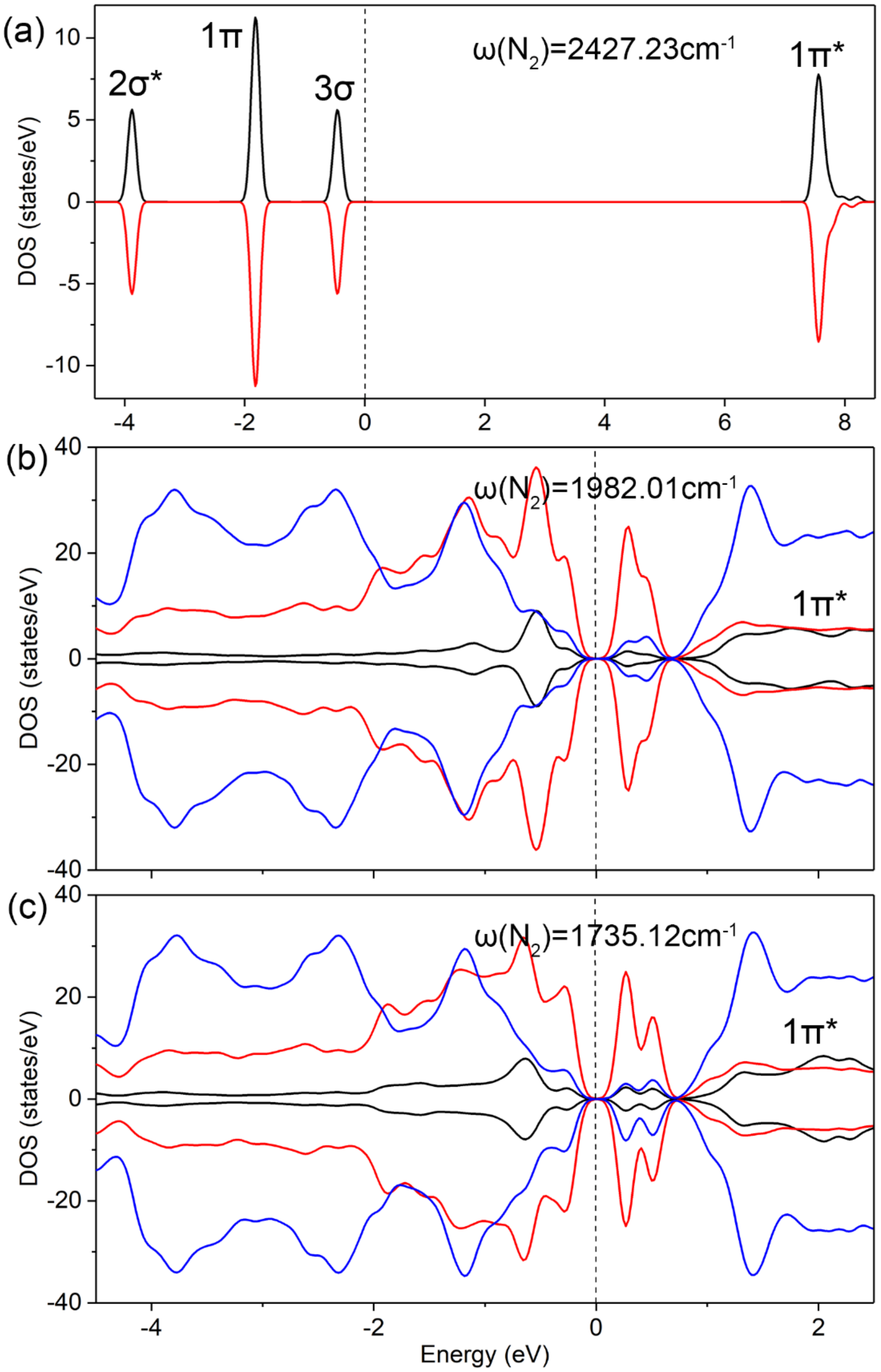

**Fig. 5** (a)The DOS of isolated $N_2$. Projected local density of states (LDOS) for (b)end-on and (c)side-on adsorption of $N_2$ above $Ru_2$@GeSe. The blue, red and black curves represent the LDOS of GeSe (blue lines), Ru dimer (red lines) and absorbed $N_2$ (black lines), respectively. Note that for helping the comparison, the values of LDOS are enlarged by a scale of 20 for absorbed $N_2$ and 10 for Ru dimer, respectively in (b) and (c). The fermi level has been set to zero.

### 3.3 The mechanism of NRR activity for $Ru_2$@GeSe

We provide a physical insight into the mechanism for the significantly high catalytic activity of $Ru_2$@GeSe towards NRR based on triple reasons given below. Firstly we calculated and analyzed the density of states (DOS) in Fig. 5 for isolated $N_2$, $N_2$ absorption on $Ru_2$@GeSe with both end-on and side-on patterns. Isolated free $N_2$ gas molecule has sharp 2σ*, 1π, 3σ and 1π* levels (Fig. 5a). The DOS diagrams show that absorbed $N_2$ have formed strong hybrid orbitals with $Ru_2$@GeSe as reflected by a significant broadening of antibonding (1π*) orbital for both Fig. 5(b) end-on and Fig. 5(c) side-on configurations. This implies that $Ru_2$@GeSe offers a strong capability to absorb and activate $N_2$ through coupling of antibonding orbitals.

Secondly, the $Ru_2$ introduces several empty levels within the band gap of GeSe (see band structure of $Ru_2$@GeSe in Fig. S3). Such defective states would enhance electrical conductivity which also equally plays important role for electrochemical NRR to achieve a fast charge transfer[53]. In contrast with the normal occupied hybridized Ru-GeSe states in valence or conduction bands, these localized diatomic states in the gap allows mixture with the evanescent state of GeSe which are normally forbidden within the band gap. Therefore, the $Ru_2$ effectively filters GeSe states which are further hybridized with the adsorbed $N_2$ molecule. As shown in Fig. 5 (b) and (c), the $N_2$ adsorbed $Ru_2$@GeSe complex contains prominent amounts of the GeSe states which are resonant with the $Ru_2$ dimer and $N_2$ molecule. It should be noted that similar resonant states are also present in the valence top of GeSe and explains the back-donation related charge transfer from GeSe to the $N_2$ molecule shown in Fig. 2 and Fig. S4.

Thirdly, we also calculated the vibrational frequencies of isolated $N_2$ and absorbed $N_2$ via displacement approach. The vibrational frequency of the stretching mode of isolated $N_2$ is 2427.23 $cm^{-1}$, whereas 1982.01 $cm^{-1}$ and 1735.12 $cm^{-1}$ are found for end-on and side-on, respectively, showing a drop of frequency up to 600 $cm^{-1}$. This significant softening trend for adsorbed $N_2$ reflects a weakened N-N bond which triggers dissociation of $N_2$ for NRR.

### 3.4 Potential photocatalytic activity and stability of $Ru_2$@GeSe for NRR

Similar to phosphorene, GeSe is well-known for its superior response to light excitation, in particularly within the near-infrared spectrum, together with a much better stability under ambient conditions than phosphorene[54,55]. This inspires the design of promising photo-driven functional materials based on GeSe. Considering the superior NRR activity, Ru-doped GeSe is promising for photocatalysis. As for photocatalysts, an important evaluation criterion is their photoconversion efficiency[10]. We calculated the absorption spectra for other typical semiconducting 2D materials including phosphorene, $MoS_2$ and g-$C_3N_4$ which are compared with GeSe. As shown in Fig. 6, there exists the strongest absorption between ~1.6 eV and ~ 3.0 eV for pristine GeSe. Interestingly, the GeSe shows a similar but slightly red-shifted absorbing spectrum compared with $MoS_2$. Both these two d-orbital contained semiconductors show a much stronger light response compared with the left p-orbital governed phosphorene and g-$C_3N_4$. Owing to its phosphorene-like structure, the optical response of GeSe is clearly anisotropic along the zigzag and armchair directions, but this anisotropy is much weaker than that of phosphorene. This quasi-isotropic response helps to promote the light utilization within certain range of energy and increase the amounts of excited carriers from different polarized light.

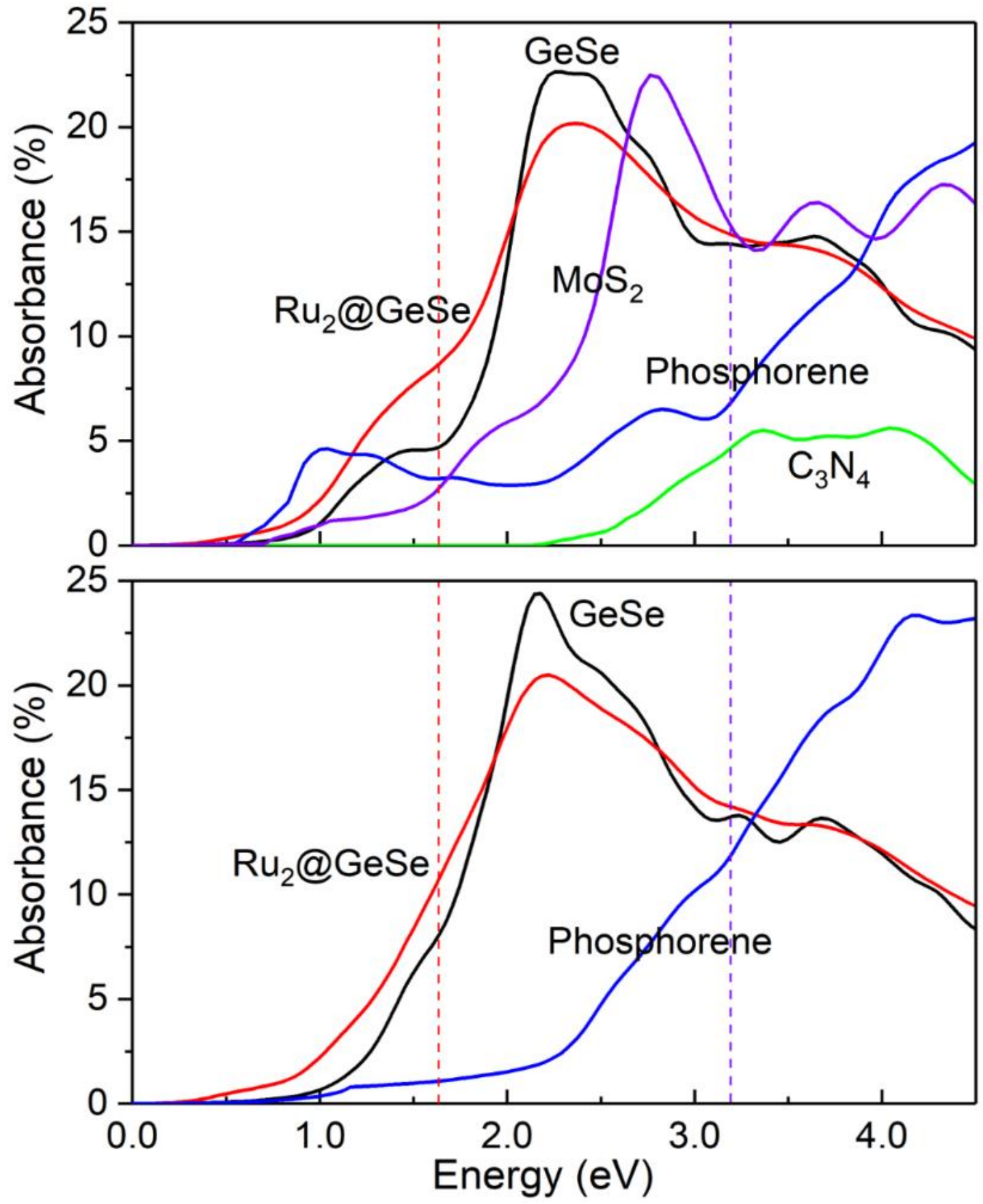

**Fig. 6** The optical absorption spectra of GeSe, $MoS_2$, phosphorene and g-$C_3N_4$ along the zigzag (top panel) and armchair (bottom panel) directions.

The addition of the Ru dimer above the GeSe broadens the window around 0.3 eV toward the low-energy branch and significantly enhances the absorption for those low-energy photons from ~1.0 to ~1.5 eV. This could be due to presence of a series of mid-gap states of Ru as shown in band structures of the $Ru_2$@GeSe in Fig. S3. Those empty levels enables additional electronic transits from the valence band states of GeSe to the Ru orbitals. Importantly, as the valence top of the GeSe at the zone center consists of significant amount of even-parity Ge-s state[56], which can principally tunnel through Ru dimer and couple with the odd-parity 1π* of the $N_2$ molecule. This might trigger a photocatalytic reduction of the $N_2$.

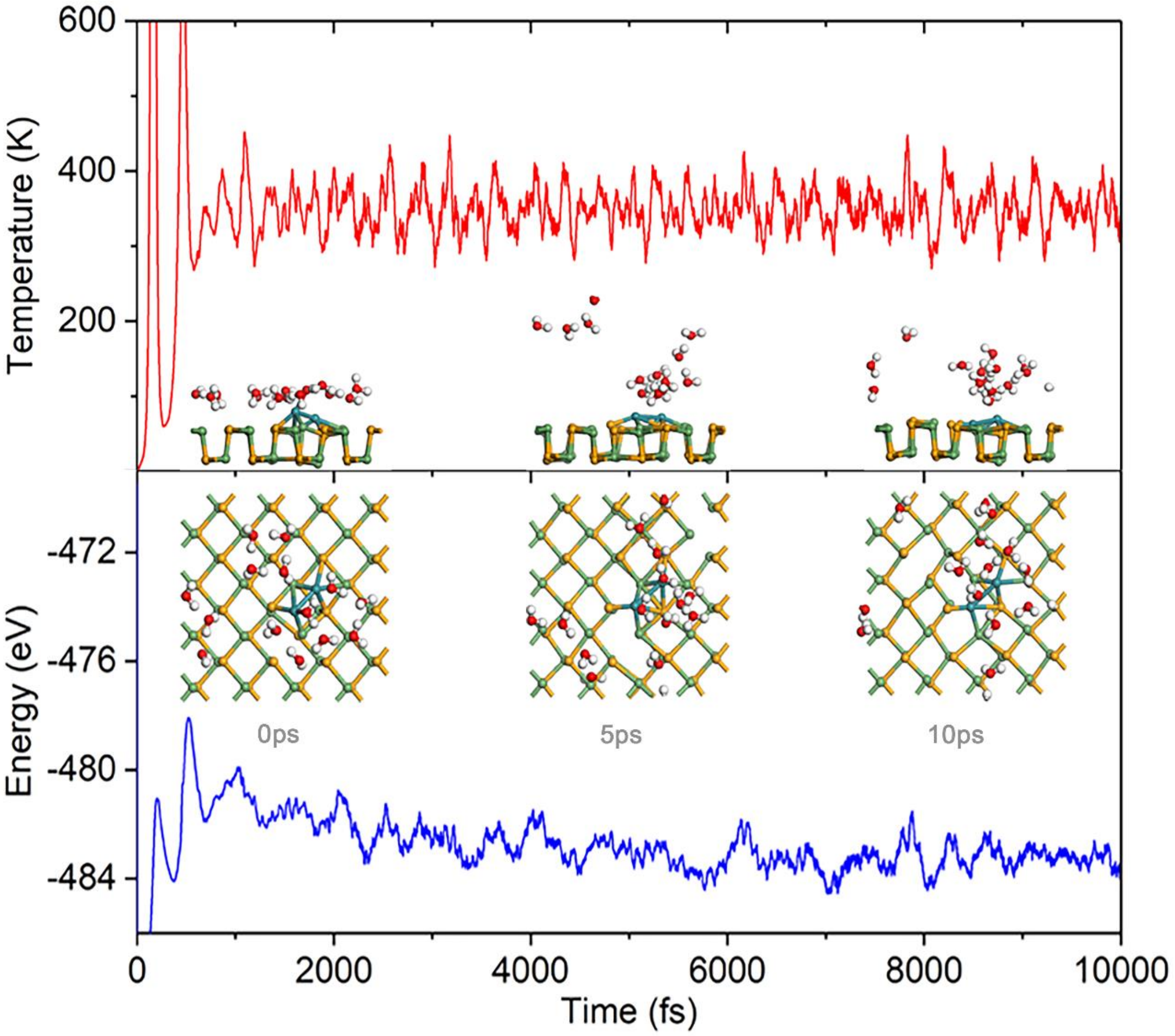

**Fig. 7** The variation of temperature and energy in the presence of water with time for the AIMD simulation of $Ru_2$@GeSe.

Finally, we examine the stability of the $Ru_2$@GeSe catalysts. In order to

demonstrate the thermodynamic stabilities of $Ru_2$@GeSe compounds, we firstly calculate the binding energies $E_b$. Our results show that the binding energy of Ru dimer decorated on GeSe monolayer is as high as -5.04 eV, indicating a strong decoration. Furthermore, we calculated the corresponding cohesive energies $E_c$ and negative ($E_b$ – $E_c$) represents high thermodynamic stability. The ($E_b$ – $E_c$) of $Ru_2$@GeSe is -1.36 eV which denotes $Ru_2$@GeSe has superior thermodynamic stability. The $E_b$ and ($E_b$ - $E_c$) of other $D_x$@GeSe systems are listed in Fig. S5. By the way, the binding energies of diatomic dimers (B, Fe, W, Mo and Ru dimers) are from diatomic dimer. Next, the thermal stability of $Ru_2$@GeSe was further confirmed via AIMD simulation. In order to simulate more accurate electrochemical reduction conditions, a water layer with 14 water molecules is formed on $Ru_2$@GeSe after structure relaxation. We simulated this system up to 10 ps with 1 fs per step under NPT ensemble. We tested the thermodynamic stability of Ru2@GeSe in the presence of water to simulate real NRR condition. As shown in Fig. 7, a good stability of the Ru dimer is found even with the presence of water, as reflected by the slight structural change (snapshots at 5ps and 10ps), robust temperature and energy fluctuations. Such stability of the Ru dimer is also reflected from the slightly fluctuated bond length of the Ru-Ru for the whole hydrogenation process as shown in Fig. 4 (c).

## 4. Conclusions

In summary, we examine the atomic scale process for the NRR above a phosphorene-like puckering surface of germanium selenide as a support for atomic catalyst. Through examining a series of typical TM and non-TM atoms via first-principles thermodynamics analysis, we find that the rippled structure facilitates the interactions between the catalysts and the support. We predict that $Ru_2$@GeSe is a promising catalyst for $N_2$ fixation and reduction based on $\Delta G_{MAX}$ as the figure of merit. The Ru dimer as the active catalytic site is predicted to have a relatively low $\Delta G_{MAX}$ down to 0.26 eV through the enzymatic pathway and 0.25 eV through the distal pathway. We reveal that the filtered states of GeSe by Ru dimer could be the underlying mechanism for the promoted activation of inert nitrogen bonds and subsequent reduction of nitrogen by hydrogenation. This strategy may pave a way to the design of potential catalysts for electrochemical nitrogen reaction reduction.

## Acknowledgement

This work is supported by the University of Macau (SRG2019-00179-IAPME) and the Science and Technology Development Fund from Macau SAR (FDCT-0163/2019/A3), the Natural Science Foundation of China (Grant 22022309) and Natural Science Foundation of Guangdong Province, China (2021A1515010024). This work was performed in part at the High Performance Computing Cluster (HPCC) which is supported by Information and Communication Technology Office (ICTO) of the

University of Macau.